\documentclass[aps,prd,reprint,groupedaddress,nofootinbib]{revtex4-1}
\pdfoutput=1

\usepackage{amsmath,amssymb}
\usepackage{graphicx}
\usepackage{color}
\usepackage{slashed}
\usepackage[english]{babel}
\usepackage[activate={true,nocompatibility},final,tracking=true,kerning=true,spacing=true,factor=1100,stretch=10,shrink=10]{microtype}

\definecolor{cadmiumgreen}{rgb}{0.0, 0.5, 0.3}

\definecolor{dgreen}{rgb}{0.5, 0.5, 0.0}

\definecolor{dblue}{rgb}{0.0,0.0,0.4} 

\usepackage{epsfig}
\usepackage[colorlinks=true,urlcolor=dblue,anchorcolor=blue,citecolor=blue,filecolor=blue,linkcolor=blue,menucolor=blue,pagecolor=blue,linktocpage=true,pdfproducer=medialab,pdfa=true]{hyperref}

\graphicspath{{./figs/}}

\def\figwidth{0.7\textwidth}

\begin{document}

\preprint{IFIC/16-01}
\title{Neutrino propagation in the galactic dark matter halo}

\author{\href{http://goo.gl/CHlgQR}{P.~F.~de~Salas}}
\affiliation{
Instituto de F\'{\i}sica Corpuscular -- CSIC/U. Valencia,\\ Parc Cient\'{\i}fic, calle 
Catedr\'{a}tico Jos\'{e} Beltr\'{a}n 2, E-46980 Paterna, Spain\\}

\author{\href{http://goo.gl/00TnL}{R.~A.~Lineros}}
\affiliation{
Instituto de F\'{\i}sica Corpuscular -- CSIC/U. Valencia,\\ Parc Cient\'{\i}fic, calle 
Catedr\'{a}tico Jos\'{e} Beltr\'{a}n 2, E-46980 Paterna, Spain\\}

\author{\href{http://goo.gl/x8CPgJ}{M.~T\'{o}rtola}}
\affiliation{
Instituto de F\'{\i}sica Corpuscular -- CSIC/U. Valencia,\\ Parc Cient\'{\i}fic, calle 
Catedr\'{a}tico Jos\'{e} Beltr\'{a}n 2, E-46980 Paterna, Spain\\}

\begin{abstract}
Neutrino oscillations are a widely observed and well established phenomenon.
It is also well known that deviations with respect to flavor conversion probabilities in vacuum arise due to neutrino interactions with matter.
In this work, we analyze the impact of new interactions between neutrinos and the dark matter present in the Milky Way on the neutrino oscillation pattern.
The dark matter-neutrino interaction is modeled by using an effective coupling proportional to the Fermi constant $G_F$ with no further restrictions on its flavor structure. 
For the galactic dark matter profile we consider an homogeneous distribution as well as several density profiles, estimating in all cases the size of the interaction required to get an observable effect at different neutrino energies. 
Our discussion is mainly focused in the PeV neutrino energy range, to be explored in observatories like IceCube and KM3NeT.
The obtained results may be interpreted in terms of a light $\mathcal{O}$(sub-eV--keV) or WIMP-like dark matter particle or  as a new interaction with a mediator of $\mathcal{O}$(sub-eV--keV) mass.
\end{abstract}


\date{\today}
\maketitle

\section{Introduction}
\label{sec:intro}

Neutrinos have been observed over a quite wide energy range, from eV to hundreds of TeV, beyond the most energetic part of the atmospheric neutrino flux.
The detection of even higher energetic neutrinos is now possible thanks to experiments like IceCube~\cite{Aartsen:2013jdh,Aartsen:2014gkd} and the future KM3NeT~\cite{Coniglione:2015aqa}, in the southern and northern hemisphere respectively. 
It is also well known that neutrinos are massive and mixed, and as a consequence they experience flavor oscillations~\cite{Agashe:2014kda}.
Moreover, it was noticed that these oscillations get modified in presence of matter due to the Mikheyev-Smirnov-Wolfenstein (MSW) effect~\cite{Wolfenstein:1977ue,Mikheev:1986gs}.
On the other hand,  a great amount of evidences indicate that most of the matter in our Universe is in the form of Dark Matter (DM), being our galaxy embedded in a DM halo~\cite{2012arXiv1201.3942P,2015NatPh..11..245I}.
From the point of view of low-energy neutrinos (below $1\,\mathrm{TeV}$), the effect of ordinary matter in neutrino oscillations through the MSW effect has been well studied.
However, if neutrinos interact with DM, their oscillations may also be affected, specially in the less explored case of ultra-high-energy neutrinos.
In this paper we are going to explore this possibility, studying the possible effect of DM on neutrino oscillations. We will focus on very-high-energy neutrinos, like the ones detected by IceCube.

Similar hypothesis have been discussed in previous works. Most of them, however, consider neutrinos as the main component of the galactic DM~\cite{Horvat:1998ym,Lunardini:2000swa} or the effect due to the interaction on the Cosmic Neutrino Background~\cite{Diaz:2015aua}. The more recent analysis in Ref.~\cite{Miranda:2013wla} follows a closer approach, although it does not account for the effect of the DM halo density profile of our galaxy.
In this work we show that the effect of the DM profile may be very important, specially for neutrinos generated near the Galactic Center (GC). This occurs because DM density variations can lead to a resonant behavior in neutrino oscillations, as it happens in the Sun~\cite{Wolfenstein:1977ue,Mikheev:1986gs}. 
A further motivation for our analysis are the recent results from the IceCube Collaboration~\cite{2015arXiv151005223T} on the flavor composition of very high-energy neutrinos.
IceCube data indicate that the preferred flavor composition of the neutrino flux is far from the expected region obtained assuming standard astrophysical neutrino sources~\cite{2015PhRvL.115p1302B,Mena:2014sja,Palomares-Ruiz:2015mka}, although the statistical significance of the discrepancy is small. 
Indeed, if we just consider neutrinos coming from neutron decay, the theoretical expectations and the observed flavor composition disagree at the $3.7\sigma$ level.
However, if the observed astrophysical neutrinos have been produced from pion decay, the tension with the experimental data is reduced to $1\sigma$. 
In any case, and provided that a better estimation of the flavor composition will be done in the near future, it is important to understand the origin of the discrepancy.
A modification in the oscillatory neutrino pattern induced by DM, for instance, can help explaining the disagreement between the predicted and observed flavor composition of the neutrino flux.
Furthermore, this hypothesis may imply new consequences in future high-energy neutrino observations. 
In particular, and given the anisotropic character of the phenomenon due to the non-central position of the solar system inside the Milky Way, the presence of galactic DM may predict different neutrino flavor composition for KM3NeT and IceCube, sensitive to different regions of the sky.\\

The origin of very-high-energy neutrinos is not completely understood. 
However, there are some estimations suggesting that, for energies larger than 60 TeV, ~40\% of the total neutrino flux has a galactic origin while the rest is extragalactic~\cite{Palladino:2016zoe}.
Even though galactic neutrinos are less abundant at these energies, the present study is focused in such component to analyze the effect of the DM halo of the Milky Way on the neutrino propagation.
The case of extragalactic neutrinos will be studied  in a future work~\cite{pfds:2016}.\\

The paper is organized as follows.
In Section~\ref{sec:dm}, we discuss the presence of DM in our galaxy.
In Section~\ref{sec:neut}, we review the main aspects of neutrino oscillations, discussing how they are affected by the presence of dark matter.
This effect is analyzed in a model independent way, parameterizing the interaction of neutrinos and dark matter in terms of an effective potential.
We show the correlations between the flavor composition of the neutrino flux at Earth and the neutrino production point.
The implications of our results are discussed in detail in Section~\ref{sec:disc}.
Finally, our conclusions are summarized in Section~\ref{sec:conc}.\\

\section{Dark matter distribution and candidates}
\label{sec:dm}
Dark matter together with dark energy are the two largest components of the energy budget of the Universe, controlling many aspects of its evolution.
Dark energy drives the accelerated expansion of the Universe while dark matter is responsible for the structure formation at different scales including the galaxy formation.
The latest  Cosmic Microwave Background (CMB) observations done by the Planck Collaboration~\cite{2015arXiv150201589P} set the present DM abundance in the $\Lambda$CDM model to
\begin{equation}
\Omega_{\rm DM}h^2 = 0.1198 \pm 0.0015 \, ,
\end{equation}
where $h = H_0/(100 \, \rm{km} \,\rm{s}^{-1}\rm{Mpc}^{-1})= 0.678 \pm 0.009$ is the scale factor for the Hubble expansion rate~\cite{2015arXiv150201589P}.\\

From observations of galaxy rotation curves~\cite{1976AJ.....81..687R,1976AJ.....81..719R} and N-body numerical simulations~\cite{1998ApJ...499L...5M,1998MNRAS.300..146G}, one may conclude that the dark matter distribution in galaxies follows a \emph{universal} profile.
For the scope of this work, we describe the DM distribution in the Milky Way by using two extreme choices: the isothermal profile~\cite{1980ApJS...44...73B} and Navarro-Frenk-White (NFW) profile~\cite{1996ApJ...462..563N,1997ApJ...490..493N}. Both parameterizations are generically described by the functional form
\begin{equation}
\rho_{\rm DM}(r,r_s,\alpha,\beta,\gamma) = \rho_{\oplus} \left(\frac{r_{\oplus}}{r}\right)^{\gamma} \, \left(\frac{1+ (r_{\oplus}/r_s)^{\alpha}}{1+ (r/r_s)^{\alpha}} \right)^{(\beta-\gamma)/\alpha} \, ,
\end{equation}
where spherical symmetry is assumed and the origin, $r = 0$, corresponds to the galactic center. 
We consider the solar system is located at $r_\oplus = 8.5~{\rm kpc}$ from the galactic center and the local DM energy density is $\rho_\oplus = 0.4~{\rm GeV}/{\rm cm}^{3}$, as indicated by several studies~\cite{2010JCAP...08..004C, 2015arXiv150406324P, 2015arXiv151006810X, 2015ApJ...814...13M}.
The isothermal DM profile is then described by the choice
\begin{equation}
\label{eq:iso}
\rho_{\rm iso}(r) = \rho_{\rm DM}(r,5~{\rm kpc},2,2,0) \, ,
\end{equation}
while the NFW profile is given by
\begin{equation}
\label{eq:nfw}
\rho_{\rm NFW}(r) = \rho_{\rm DM}(r,20~{\rm kpc},1,3,1) \, ,
\end{equation}
where the main difference between both distributions is the presence of a cusp at $r=0$ in the NFW profile.\\

However, more than the issue of the DM distribution inside our galaxy, the key point to deal with DM is its true nature, what constitutes one of the biggest puzzles in physics nowadays.
The most popular candidates to DM are generically known as Weakly Interactive Massive Particles (WIMPs) \cite{2012arXiv1201.3942P,Steigman:1984ac, 2012arXiv1204.2373D} and they appear in many models beyond the Standard Model (SM).
The WIMP relic abundance arises from the epoch when  WIMPs and SM particles were in thermal equilibrium.
Due to the expansion of the Universe, WIMPs cannot remain in thermal equilibrium, \emph{freezing out} the WIMP abundance and forming the current DM abundance.
In the context of the thermal freeze-out mechanism, the observed relic abundance is then obtained when the thermally averaged WIMP cross section is $\langle \sigma v \rangle \simeq 3 \times 10^{-26} \, {\rm cm}^3/{\rm s}$ at the freeze-out temperature $T_{\rm f.o.} \simeq m_{\rm WIMP}/20$.
This mechanism leads to a WIMP DM mass in the range of GeV--TeV and sets the order of magnitude of the DM interaction with the SM particles.
This class of candidates opens the possibility to look for DM annihilation  at celestial objects (indirect detection) as well as 
DM recoils on super sensitive detectors located at underground laboratories (direct detection).\\

Asymmetric DM~\cite{2013IJMPA..2830028P,2014PhR...537...91Z} is another popular candidate. 
In this case, the DM abundance is generated via (model-dependent) mechanisms similar to baryogenesis~\cite{2013IJMPA..2830028P} leading to a DM mass of the order of GeV.
The main implication arising from this scenario is a DM content of the Universe made of DM particles without any DM antiparticle.
In consequence, there are no DM annihilations to be detected through indirect detection signals.
Direct searches are still possible due to the interaction of DM with normal matter.\\

Possible candidates to DM particles include also sterile neutrinos, majorons and axions.
Sterile neutrinos and majorons may be connected with the generation of neutrino masses through different mechanisms~\cite{2014NJPh...16l5012L} while axions are related to the strong CP problem~\cite{2009NJPh...11j5008D}.
In general, all these candidates have their mass in the sub-MeV range although one can find candidates with even lighter mass~\cite{Hu:2000ke}.
Note that the zoology of DM models is quite broad~\cite{Zhang:2015ffa, Suarez:2013iw} and the ones discussed above provide  a  brief sample of  the most popular candidates.
\\

As summarized above, the masses of most of the proposed DM candidates can range from eV to TeV.
For the sake of simplicity, along this work we will assume  a generic value for the mass of the DM particle as well as a generic value for its coupling to neutrinos.
In the discussion of our results in the next section we will consider three different benchmark cases with particular values for these parameters.\\

\section{Neutrino oscillations in dark matter}
\label{sec:neut}

Neutrinos coming from the Milky Way, from sources outside the solar system, may be affected by the DM halo of the galaxy.
In particular, their flavor oscillations might be modified due to the presence of DM in analogy with the MSW effect in ordinary matter~\cite{Wolfenstein:1977ue,Mikheev:1986gs}.
Since the nature of DM is still unknown, one can assume the most general case where its interaction with neutrinos can violate flavor universality.
This can give rise to a sizable effect when compared to neutrino oscillations in vacuum.\\

It is well established that neutrino oscillations happen because mass and weak eigenstates do not coincide:
\begin{equation}
|\nu_{\alpha} \rangle = \sum_{k} U^{*}_{\alpha k} | \nu_{k} \rangle \, ,
\end{equation} 
where  $\alpha$ is a flavor index (e, $\mu$, $\tau$), $k$ refers to neutrino mass eigenstates with mass $m_k$, and $U$ is the unitary matrix which diagonalizes the neutrino mass matrix.\\

When neutrinos travel across a medium like the Earth or the Sun, their interaction with the medium modifies the vacuum oscillation pattern.
This effect introduces a new term in the total hamiltonian describing 
the  evolution of neutrinos in the medium
\begin{equation}
\label{eq:htot}
\mathcal{H}_{\rm tot} = \mathcal{H}_{\rm vac} + \mathcal{V}\,,
\end{equation}
where $\mathcal{H}_{\rm vac}$ is the hamiltonian in vacuum in the flavor basis
\begin{equation}
\mathcal{H}_{\rm vac} = \frac{1}{2 E} \, U \left( 
\begin{array}{ccc}
0 & 0 & 0 \\
0 & \Delta m^2_{21} & 0 \\
0 & 0 & \Delta m^2_{31}
\end{array}
\right) U^{\dagger}  \, .
\end{equation}
Here  $\Delta m^2_{ij} = m^2_i - m^2_j$ are the mass squared differences between the neutrino mass eigenstates $\nu_i$ and $\nu_j$ resulting after the substraction of the global phase $m_1^2$.
$U$ is the neutrino mixing matrix~\cite{Agashe:2014kda} and $\mathcal{V}$ is the effective potential that accounts for the neutrino interactions with matter. 
For the neutrino mixing angles and mass squared differences we use  the best fit values in Ref.~\cite{2014PhRvD..90i3006F} for normal and inverse hierarchy. The CP violation phase $\delta$ has been set to zero.\\

The effective potential describing the interaction between neutrinos and DM in the flavor basis can be parametrized as
\begin{equation}
\label{eq:effpot}
\mathcal{V}_{\alpha \beta} = \lambda_{\alpha \beta} \, G_F \, N_{\chi}  \, ,
\end{equation}
where $\lambda_{\alpha \beta}$ is a hermitian matrix  encoding the effective couplings between neutrinos and DM, $G_F$ is the Fermi constant, and $N_{\chi}$ is the DM number density.
The number density is related to the energy density by
\begin{equation}
N_{\chi} = \frac{\rho_{\rm DM}}{m_{\rm DM}}\, .
\end{equation}

\begin{figure}[t]
\centering
\includegraphics[width=0.95\columnwidth]{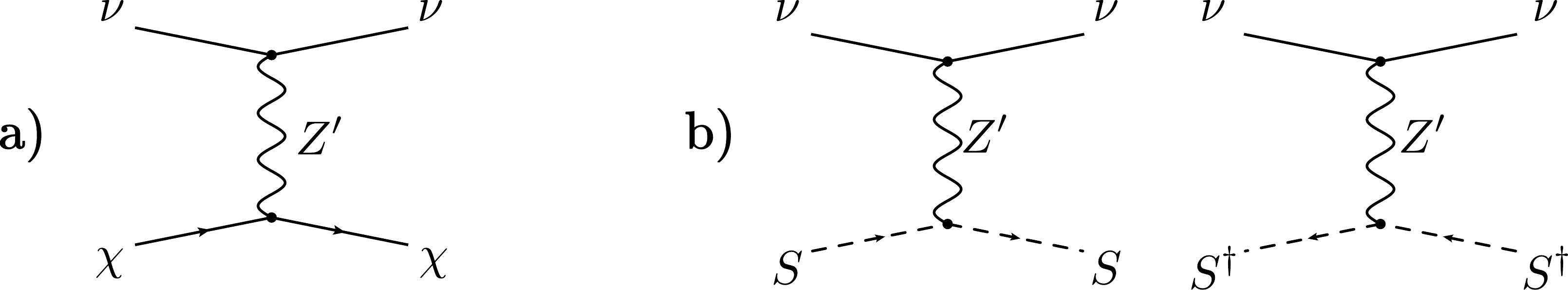}
\caption{\label{fig:zpdiag} Examples of processes leading to the parameterization in Eq.~(\ref{eq:effpot}). Diagram a) corresponds to the case of asymmetric DM while diagrams in  b) represent the case of scalar DM.}
\end{figure}

The parameterization used in Eq.~(\ref{eq:effpot}) is well motivated in scenarios with fermion asymmetric DM, where DM (and no anti-DM) is present in the Universe, as well as in the case of scalar/vector DM candidates.
Possible realizations are presented in Fig.~\ref{fig:zpdiag}.
In the simplest scenario, we require a mediator particle that connects the DM and neutrino sectors without  mixing them.
This is shown in diagram a) for a fermion asymmetric DM. 
\\

Since neutrino oscillations are blind to global phases,  we can reduce the number of free parameters in the effective potential by subtracting a term proportional to the identity matrix.
In our case we subtract $\mathcal{V}_{\tau \tau} \, \mathbb{I}$. We also consider only real entries in the $\lambda$-matrix.
Then, the effective potential is reduced to
\begin{equation}
\label{eq:effpot36}
V = \mathcal{V} - \mathcal{V}_{\tau \tau} \mathbb{I} = G_F N_{\chi} \left( 
\begin{array}{ccc}
\lambda_{11} & \lambda_{12} & \lambda_{13} \\
\lambda_{12} & \lambda_{22} & \lambda_{23} \\
\lambda_{13} & \lambda_{23} & 0 
\end{array}
\right) \, ,
\end{equation}
with five free parameters describing the effective interactions between DM and neutrinos.
Let us remark that a relative sign may appear in $\mathcal{V}$ modelling the effect for neutrinos and antineutrinos, although this sign would depend on the nature of DM candidate.
In our study, we consider all possible signs in the entries of the potential to cover the DM effect including both species without distinction.
This is motivated by the fact that neutrino telescopes are unable to distinguish between neutrinos and antineutrinos. Even though there are small differences in the $\nu / \bar{\nu}$ cross sections with the target material~\cite{Formaggio:2013kya,Adrian-Martinez:2016fdl}, the implications of such differences at the detection level are beyond the scope of this work.\\

The evolution equation of neutrinos passing through a medium is given by
\begin{equation}
i\frac{\partial\Psi}{\partial t} = \mathcal{H}_{\rm tot} \Psi,
\end{equation}
where $\mathcal{H}_{\rm tot}$ is the hamiltonian in the flavor basis as given in Eq.~(\ref{eq:htot}) and $\Psi$ the neutrino field.
The hamiltonian in the mass basis will be obtained after rotating $\mathcal{H}_{\rm tot}$ with the neutrino mixing matrix $U$,
\begin{equation}
\mathcal{H}_{\mathrm{tot}}^m  = U^\dagger \mathcal{H}_{\mathrm{tot}} U = \frac{1}{2 E}
\left( \begin{array}{ccc}
0 & 0 & 0 \\
0 & \Delta m_{21,\mathrm{eff}}^2 & 0 \\
0 & 0 & \Delta m_{31,\mathrm{eff}}^2
\end{array} \right) \, ,
\end{equation}
where $\Delta m_{21,\mathrm{eff}}^2$, and $\Delta m_{31,\mathrm{eff}}^2$ are the effective mass squared splittings in the presence of an effective potential.\\

In the case of a constant effective potential and starting with an initial flavor content $f^0 = \left( f^0_e, f^0_\mu, f^0_\tau \right)$, the final averaged flavor state will be given by
\begin{equation}
f_\beta = \sum_{\alpha = e,\mu , \tau} \left( \sum_{i=1}^{3}\left| U_{\beta i} U^*_{\alpha i}\right|^2 f^0_\alpha \right) \, ,
\label{eq:fscte} 
\end{equation}
which is valid for distances much larger than the characteristic oscillation wavelength.
%


\begin{figure*}[!t]
\centering
\includegraphics[width=0.9\textwidth]{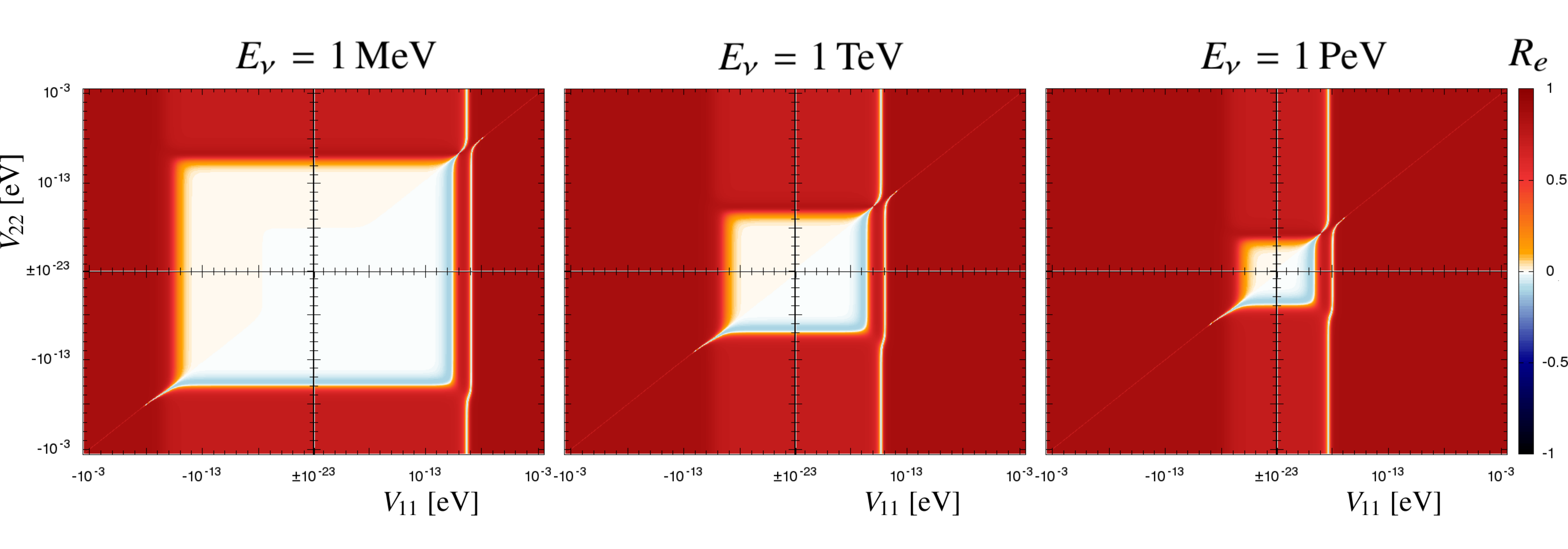}
\caption{\label{fig:boxes}   $R_e$ map in the plane $V_{11}$ -- $V_{22}$ 
starting with a flavor content (1:0:0) for $E_\nu = 1\, {\rm MeV}$ (top left panel),  $E_\nu = 1\, {\rm TeV}$ (top right panel) and $E_\nu = 1\, {\rm PeV}$ (bottom panel). }
\end{figure*}


In a more realistic scenario, the effective potential will depend on the neutrino position.
In our case, the spatial dependence arises from the DM distribution that modifies $N_{\chi}$.
The solution to the flavor evolution equation requires the diagonalization of the hamiltonian in matter at every instant, depending on the value of $N_{\chi}$ at the neutrino position.
We describe that position in terms of the line of sight distance, $l$, with respect to the solar system and the angle, $\phi$, with respect to the galactic center.
The galactocentric radius $r$ is then simply described by
\begin{equation}
r^2 = r_{\oplus}^2 +l^2 - 2 \, l\,r_{\oplus} \,\cos{\phi} \, .
\end{equation}

In this scheme, the evolution of the flavor states is given by
\begin{align}
f_{\beta}&(l_{n+1},\phi) = \nonumber \\ &\sum_{\alpha = e,\mu , \tau} \left( \sum_{i=1}^{3}\left| U_{\beta i}(l_n,\phi) U^*_{\alpha i}(l_n,\phi)\right|^2 f_\alpha(l_n,\phi) \right),
\end{align}
where the initial state corresponds to $f_\alpha(l_0,\phi) = f^0_\alpha$ with $l_0$ being the distance to the source.
 $U(l_n,\phi)$ is the matrix that diagonalizes the hamiltonian $\mathcal{H}_{\rm tot}$
evaluated at the neutrino position $(l_n,\phi)$,  and $f_\beta(l=0,\phi) = f^\oplus_\beta$ is the final state at Earth.
The distances involved in these scenarios are of the order of 1 kpc, while the largest oscillation wavelength is much smaller for neutrino energies of TeV--PeV.
Therefore, we can safely assume that the neutrino flavor evolution may be described by an averaged flavor oscillation.

On the other hand, if neutrino propagation is adiabatic,  the numerical integration of the evolution described above can be further simplified. In that case, the neutrino flavor evolution would only depend on the initial and final DM densities. As we have checked in Appendix \ref{se:adia}, for the DM densities analyzed the adiabaticity is satisfied in the neutrino propagation so we can safely calculate the final neutrino flavor composition under this assumption. We have numerically checked that this is the case.

In what follows, we will consider two different approaches to evaluate the effect of neutrino interactions with DM on the observed signal at experiments.
As a first approximation, we will consider the case of a homogeneous DM distribution.
Next, we will proceed including the more realistic case of a varying DM density profile assuming either a NFW or an isothermal DM profile.

\subsection{Case I: homogeneous DM halo}
\label{ssec:cte}

\begin{figure*}[!t]
\centering
\includegraphics[width=0.9\textwidth]{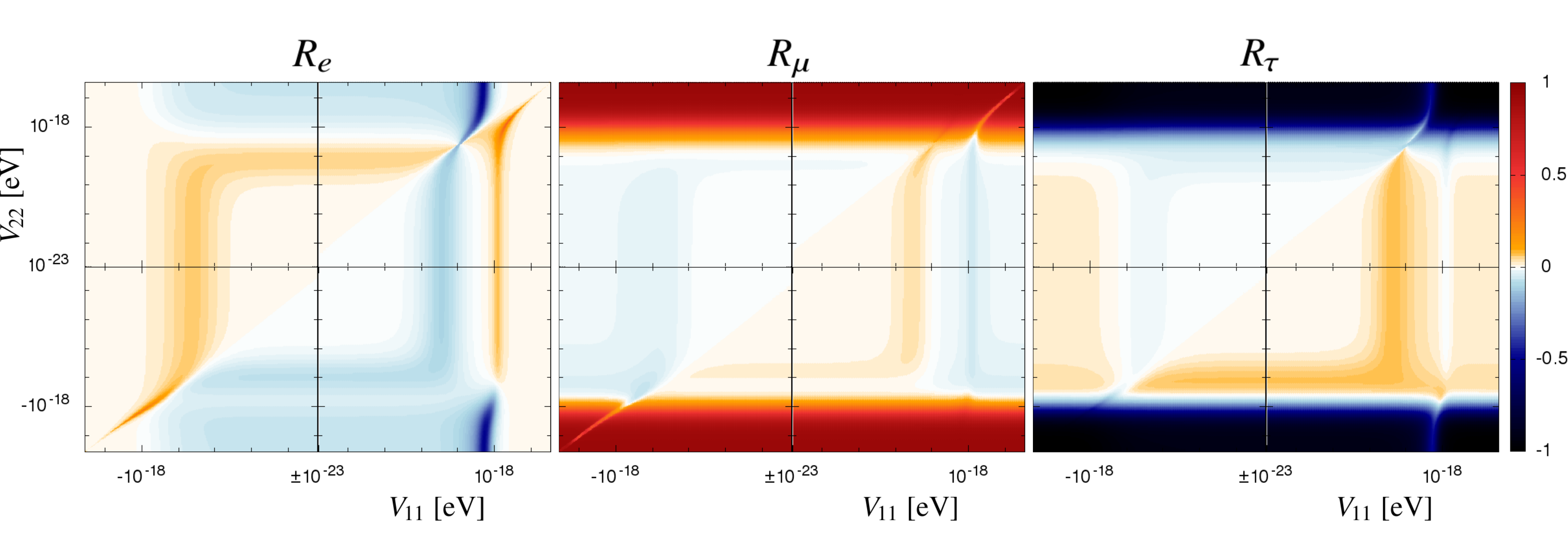}
\caption{\label{fig:1PEV} $R_\beta$ ($\beta$ = e, $\mu$, $\tau$) maps in the plane $V_{11}$ -- $V_{22}$  for $E_{\nu} = 1 \,{\rm PeV}$ and an initial flavor content of (1:2:0). 
}
\end{figure*}

In this section we will  search for values of the effective potential $V$  leading to measurable effects on the neutrino oscillation pattern.
As a first approximation to the problem, we will focus on the effects of a homogeneous DM halo.
This is equivalent to use a constant value of the effective potential and therefore the final flavor state is obtained from Eq.~(\ref{eq:fscte}).
The deviations between the DM-modified final state, $f_{\beta}^{\rm DM}$, and the expected final state in vacuum, $f_{\beta}^{\rm vac}$, at Earth  are then described by
\begin{equation}
R_\beta(V, E)= \frac{f_{\beta}^{\rm DM} - f_{\beta}^{\rm vac} }{f_{\beta}^{\rm vac}} \, , 
\label{eq:dev}
\end{equation}
where the effective potential $V$  is given at Eq.~(\ref{eq:effpot36}), $E$ is the neutrino energy and $\beta = (e,\mu,\tau)$ is the neutrino flavor.

We calculate the deviations in the neutrino flavor content for different configurations of the effective potential (i.e. varying different entries of the potential $V_{ij}$) and we find that their effect qualitatively shows similar results.
Therefore, for simplicity,  we consider that only the entries $V_{11}$ and $V_{22}$ are different from zero.
We then  found regions where $V_{ij}$ gives rise to strong modifications in the behavior of neutrino oscillations, with abrupt changes in the sign of $\partial R/\partial V_{ij}$.
These regions are a clear hint for a resonant behavior at the neutrino oscillations.
The transition can be observed in Fig.~\ref{fig:boxes}, where the sudden changes in color are related to the presence of a resonance.

Due to the nature of neutrino oscillations in a medium, an increase in $E_{\nu}$ would be equivalent to an increase in $V_{ij}$.
Therefore, the effects of DM on neutrino oscillations will become more relevant at higher neutrino energies.
In addition, the ratio $R_\beta$ in Eq.~(\ref{eq:dev}) reaches a saturation value when $E_\nu$ grows and $V_{ij}$ is fixed.
In Fig.~\ref{fig:boxes}, we present the values of $R_e$ in the plane $V_{11}$ -- $V_{22}$ starting from the flavor configuration $f^0$=(1:0:0) at the source.
Comparing the three panels in the figure, we observe a linear change in the size of the regions due to the linear dependence on $E_\nu$.
The resonant behavior of $R_e$ around $V_{11} = 10^{-10}\,{\rm eV}$ for $E=1\,{\rm MeV}$ is shifted to $V_{11} = 10^{-19}\,{\rm eV}$ for $E=1\,{\rm PeV}$.
This resonant pattern and its shift in energies also manifest in $R_\mu$ and $R_\tau$.
In consequence, the DM-neutrino interaction could explain a non-standard flavor composition in the high-energy neutrino flux observed at Earth, as the case of recent IceCube data~\cite{2015arXiv151005223T} without compromising lower energy observations.

We also study the case of an astrophysical source of neutrinos with initial flavor content equal to (1:2:0).
In Fig.~\ref{fig:1PEV}, we present the $R_{e,\mu,\tau}$ maps in the plane $V_{11}-V_{22}$ for $E_{\nu}=1\,\mathrm{PeV}$, an energy value relevant for IceCube.
We have explored the effective potential $V_{ii}$  in the range from $10^{-23}$ to $10^{-16}\,{\rm eV}$.
In our scan, the lower limit of $V_{ii}$ does not produce appreciable deviations from the vacuum solution (i.e $R_{\beta} \simeq 0$), while the upper limit saturates the flavor oscillations beyond the resonance and then $R_{\beta}$ does not change for larger values of $V_{ii}$.

At this level, we can compare the patterns shown in Figs.~\ref{fig:boxes} and~\ref{fig:1PEV}.
To understand the difference we need to consider the definition of $R_\beta$ in Eq.~(\ref{eq:dev}).
The averaged oscillations in vacuum for an initial flavor content of  \mbox{(1:2:0)} lead to a final flavor composition
\begin{equation}
f^{\rm vac}=(0.331:0.347:0.322) \,.
\label{eq:fvac-120}
\end{equation}
In the region where both $\left|V_{11}\right|$ and $\left|V_{22}\right|$ are larger than $10^{-18}\,{\rm eV}$, the initial flavor content remains unchanged.
This is because the effective potential dominates the neutrino hamiltonian and then flavor oscillations are suppressed.
For an initial flavor composition of (1:2:0), we obtain
\begin{equation}
f^{\rm DM} = (0.333:0.667:0) \,
\end{equation}
 in the presence of DM, which is equivalent to
\begin{equation}
(R_e: R_\mu : R_\tau ) \approx (6\times 10^{-3}:0.92: -1)\, .
\end{equation}
This explains the values of the deviations $R_\beta$ at the corners of the three panels in Fig.~\ref{fig:1PEV}.

In the region where $\left|V_{11}\right|\ge10^{-18}\,{\rm eV}$ and $\left|V_{22}\right|\le10^{-18}\,{\rm eV}$, only  electron neutrino oscillations are suppressed, leaving $f^{\rm DM}_e \simeq f^0_e$ (i.e. $R_e \simeq 0$).
However, oscillations for muon and tau neutrinos are still active. 
Since the atmospheric mixing angle $\theta_{23}$ in vacuum is almost maximal, the initial content of muon neutrinos, $f_\mu^{0}=0.667$, is equally distributed between muon and tau neutrinos.
This leads to $f^{\rm DM} \approx (0.33:0.33:0.33)$, what explains the slightly negative value of $R_\mu$ and the slightly positive one of $R_\tau$ and $R_e$ obtained after the comparison with the vacuum expectations in Eq.~(\ref{eq:fvac-120}).

The final saturation region, where $V_{11}$ is small and $V_{22}$ is large, prevents muon neutrinos to oscillate.
The vacuum oscillations between electron and tau neutrinos are mainly controlled by the reactor mixing angle $\theta_{13} \simeq 9^0$ .
This implies that just a small part of the initial electron neutrino content, $f_e^{0}=0.333$, will be transferred to tau neutrinos.
Consequently $R_e$ remains small, but negative, while $R_\tau$ will have a negative value close to $-1$.\\

\begin{figure*}[!t]
\centering
\includegraphics[width=0.45\textwidth]{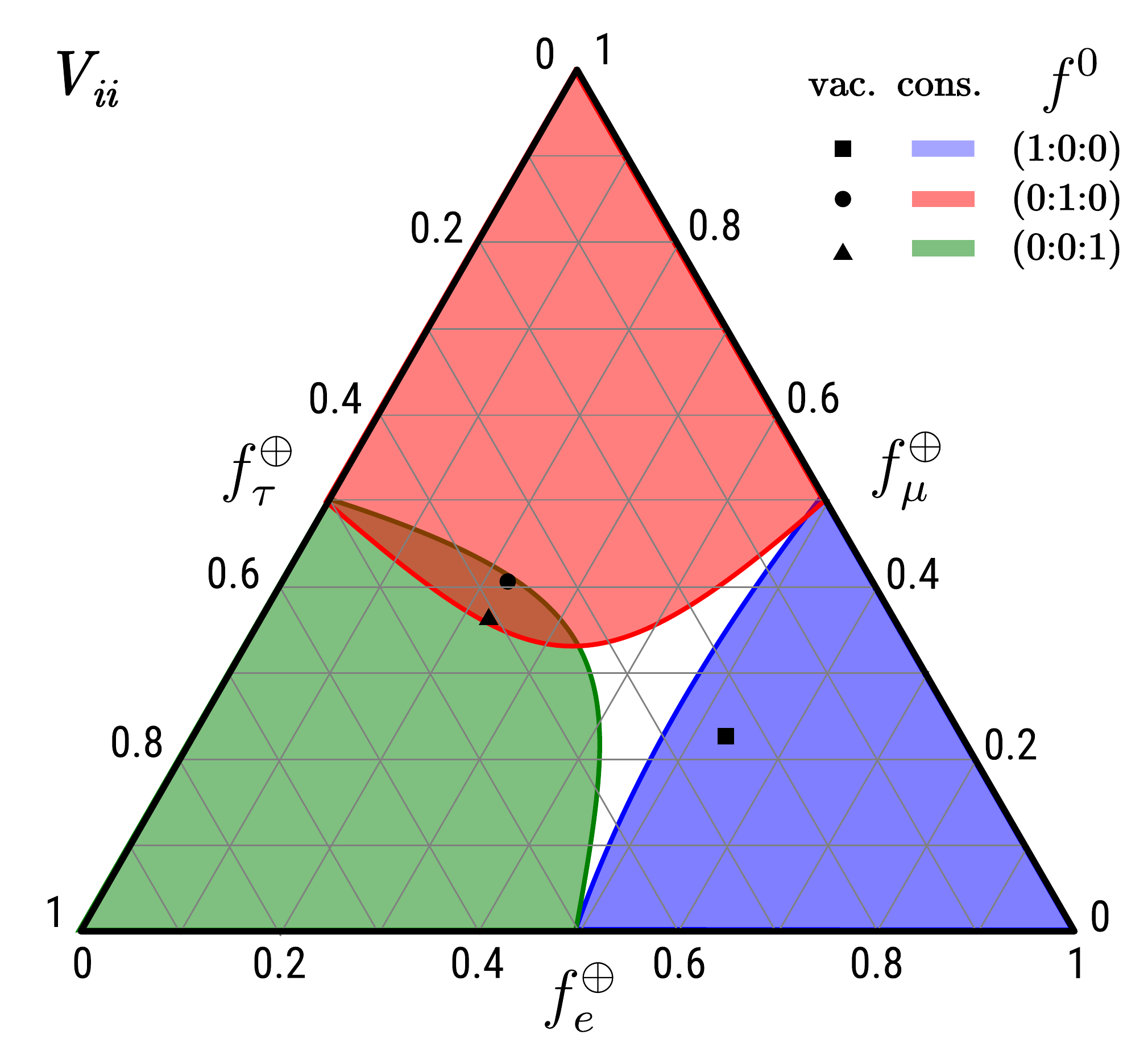}
\includegraphics[width=0.45\textwidth]{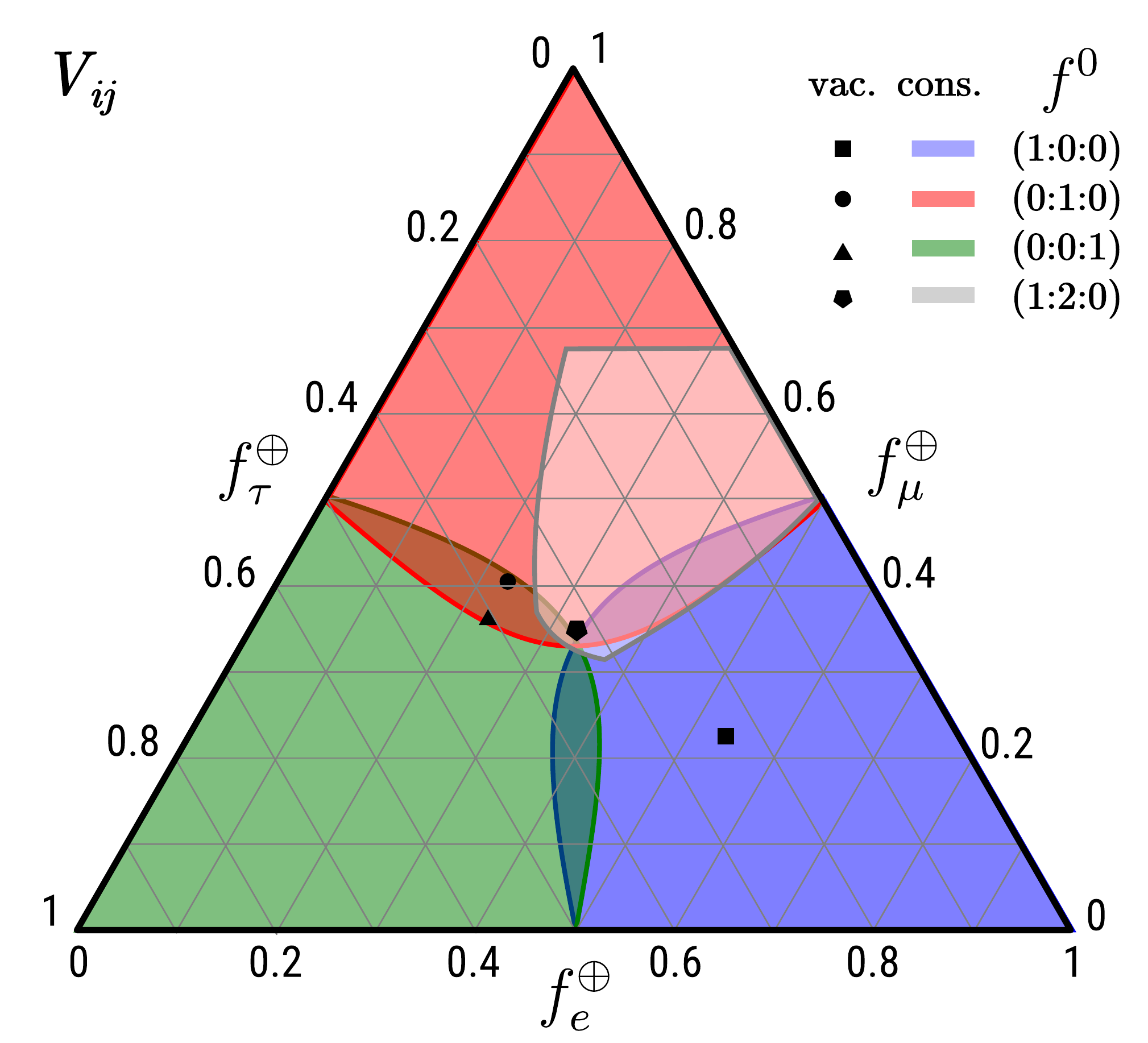}
\caption{Flavor triangle for $E_{\nu} = 1$~PeV for the initial neutrino compositions (1:0:0), (0:1:0), (0:0:1) and (1:2:0).
Color regions include the possible flavor neutrino compositions at Earth predicted in the presence of DM, whereas the points correspond to the expected flavor composition in vacuum.
 Left panel results  from the scan using only pure initial neutrino states and diagonal terms in the potential while the right panel has been obtained scanning over all entries, including the non-diagonal ones.}
\label{fig:triancte}
\end{figure*}

A more general analysis, where all the entries of the effective potential $V_{ij}$ are free to vary, is summarized in Fig.~\ref{fig:triancte}.
The regions in color cover all the possible flavor neutrino compositions at Earth predicted in the presence of DM.
In the same figure, the three points indicate the predictions assuming neutrino oscillations in vacuum.
We present our results for four different initial states: (1:0:0), (0:1:0), (1:2:0) and (0:0:1). 
The first three cases are motivated by astrophysical processes as neutron decay, damped muon source and pion decay, respectively. The last case, (0:0:1), is shown for comparison. A compendium of possible astrophysical neutrino sources are described in Ref.~\cite{2015PhRvL.115p1302B}.
In the left panel, we only vary the diagonal terms of the effective potential keeping the non-diagonal equal to zero, while in the right panel, all terms are allowed to vary, showing the maximum deviation area.
In both cases, $\left|V_{ij}\right|$ take values in the range $10^{-23}$ -- $10^{-13}\,{\rm eV}$.
From the figure we see that there is no difference between the regions obtained for  (0:1:0) and (0:0:1)  in the two panels.
On the contrary, for the initial flavor content (1:0:0), the diagonal-only scan can not cover all the region obtained in the right panel. 
Our results agree with the analysis presented in Ref.~\cite{Arguelles:2015dca} which considers possible New Physics scenarios.\\

In this part, we have analyzed the order of magnitude of the effective potential which may give a sizable deviation from the neutrino oscillation pattern in vacuum for an homogeneous DM profile.
However, one may expect that the effective potential will take different values depending on the neutrino position with respect to the DM halo.
This more realistic scenario will be analyzed in the next subsection.\\

\subsection{Case II: DM halo profile}
\label{ssec:oscdist}

\begin{table}[b]
\centering
\begin{tabular}{|c |c c c|}
\hline
\phantom{xx}Benchmark\phantom{xx} & \phantom{xx}Case A\phantom{xx} & \phantom{xx}Case B\phantom{xx} & \phantom{xx}Case C\phantom{xx}\\
\hline
$V_{11}^{\oplus}$ [$10^{-21}$ eV] & 4   &  20  & 40  \\
\hline
\end{tabular}
\caption{\label{tab:bench} Benchmark cases A, B and C, defined by the effective potential value at Earth.}
\end{table}

\begin{figure*}[t]
\centering
\includegraphics[width=\figwidth]{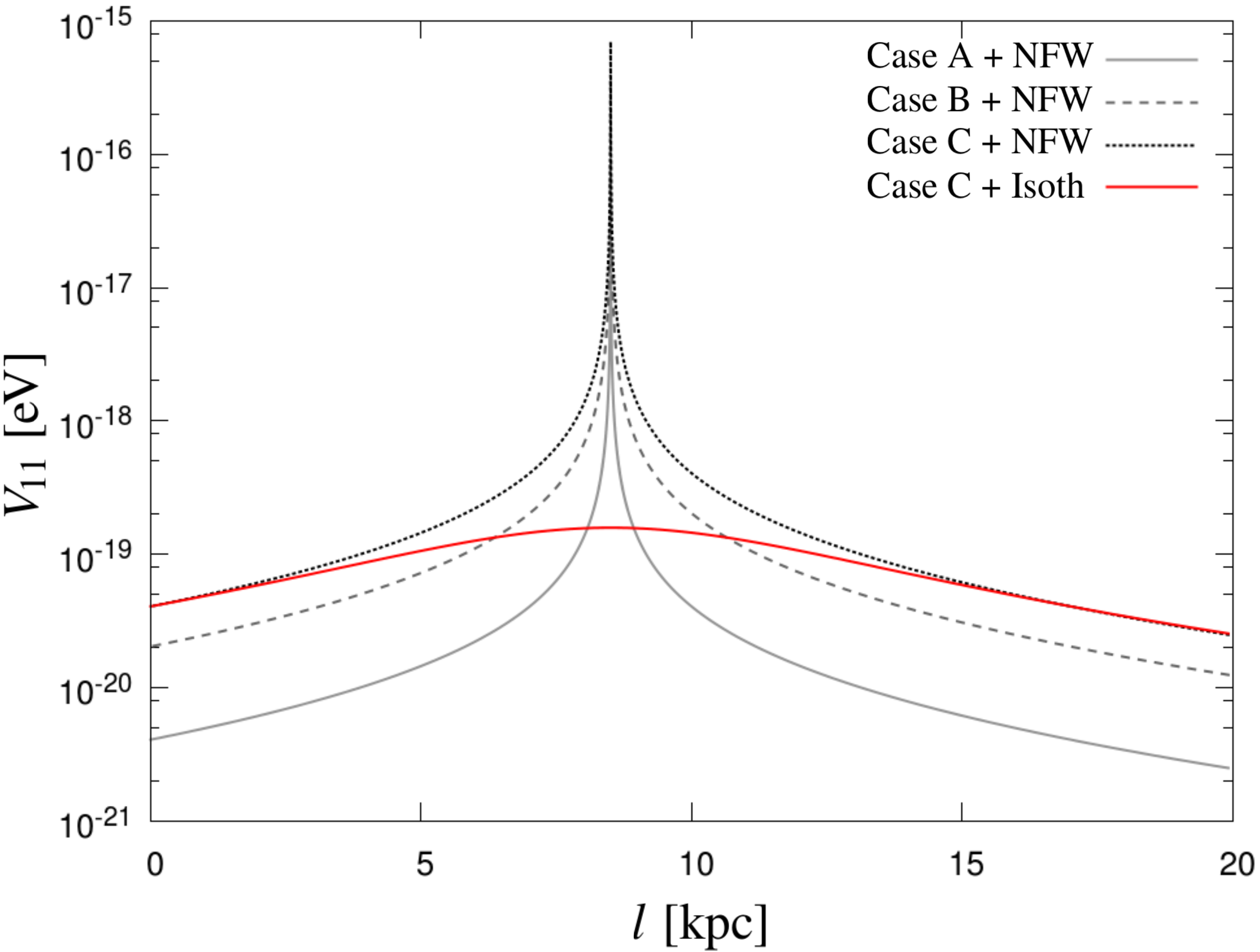}
\caption{\label{fig:case} Effective potential $V_{11}$ as a function of the distance from the Earth, $l$, for a path crossing the galactic center. Different curves correspond to the benchmark cases A, B and C in Tab.~\ref{tab:bench} for a NFW DM profile. The benchmark case C with an isothermal profile is also shown.}
\end{figure*}

In this subsection we will consider realistic profiles for the galactic DM density distribution. 
This will imply a spacial dependence in the effective matter potential along the neutrino path that may produce similar effects to the ones explaining the solar neutrino problem.
In section~\ref{ssec:cte}, we estimated the values of $V_{ij}$ for which the neutrino flavor composition at Earth is different from the case of oscillations in vacuum.
Here we combine that information with the spatial distribution of DM in the Milky Way assuming homogeneously produced neutrinos up to $r_{\rm max} = 20$~kpc on the galactic plane.
We consider three benchmark cases: A, B and C (see Table~\ref{tab:bench} for details).
The values of each benchmark are chosen to give a very small value of $R_{\beta}$ in the homogeneous DM case.
However, thanks to the DM distribution profile, the effective potential for neutrinos is larger in regions closer to the galactic center compared to the outskirts of the galaxy.
In Fig.~\ref{fig:case} we show how the effective potential changes with the DM distribution.
We consider the isothermal and NFW profiles described in Eqs.~(\ref{eq:iso}) and (\ref{eq:nfw}) to analyze the impact of the DM distribution on neutrino oscillations.
In addition, the spatial variation of the potential would produce resonances in the neutrino flavor conversion enhancing the deviations with respect to the oscillation pattern in vacuum.
%

\begin{figure*}[!t]
\centering
\includegraphics[width=0.9\textwidth]{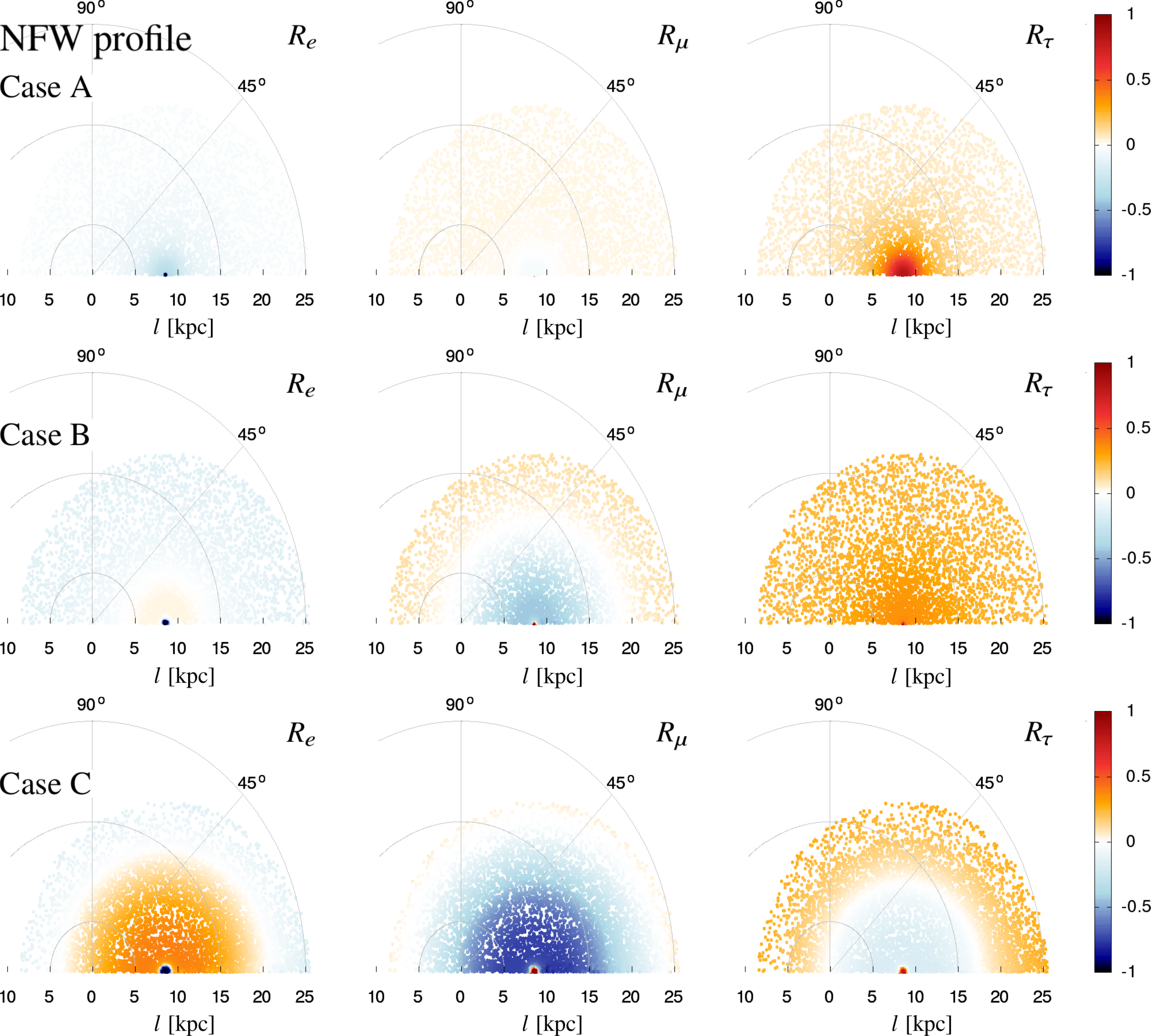}
\caption{\label{fig:circles} $R_\beta$ maps for different neutrino production position. $l = 0$ corresponds to the Earth's position. In all plots, the initial flavor state is (1:0:0), $E_\nu = 1\, {\rm PeV}$ and a NFW profile. Columns from left to right correspond to $R_e$, $R_\mu$ and $R_\tau$ respectively, and rows from top to bottom are benchmark cases A, B and C.}
\end{figure*}

In Fig.~\ref{fig:circles} we show the color map of the deviation parameter $R_{\beta}$ for an initial state (1:0:0) in terms of the neutrino production point in the galactic plane for  the NFW profile.
We consider neutrinos with energies of 1~PeV and the benchmark cases in Table~\ref{tab:bench}. 
For simplicity, only $V_{11}^\oplus$, the 11-entry of the effective potential resulting from the dark matter-neutrino interaction at Earth $V_{ij}(r_{\oplus})$,  has been chosen different from zero.
Each column of plots, from left to right, represents $R_{e}$, $R_{\mu}$ and $R_{\tau}$, respectively,
while each row corresponds to a different benchmark point.
We observe that the impact of the spatial dependence is very important and the effect is different for each $R_{\beta}$. 
The use of different benchmarks has also an impact on the $R_{\beta}$ maps. 
As expected, one can notice that the benchmark C presents the largest deviations over every $R_{\beta}$ if compared with benchmarks A and B.

\begin{figure*}[t]
\centering
\includegraphics[width=0.9\textwidth]{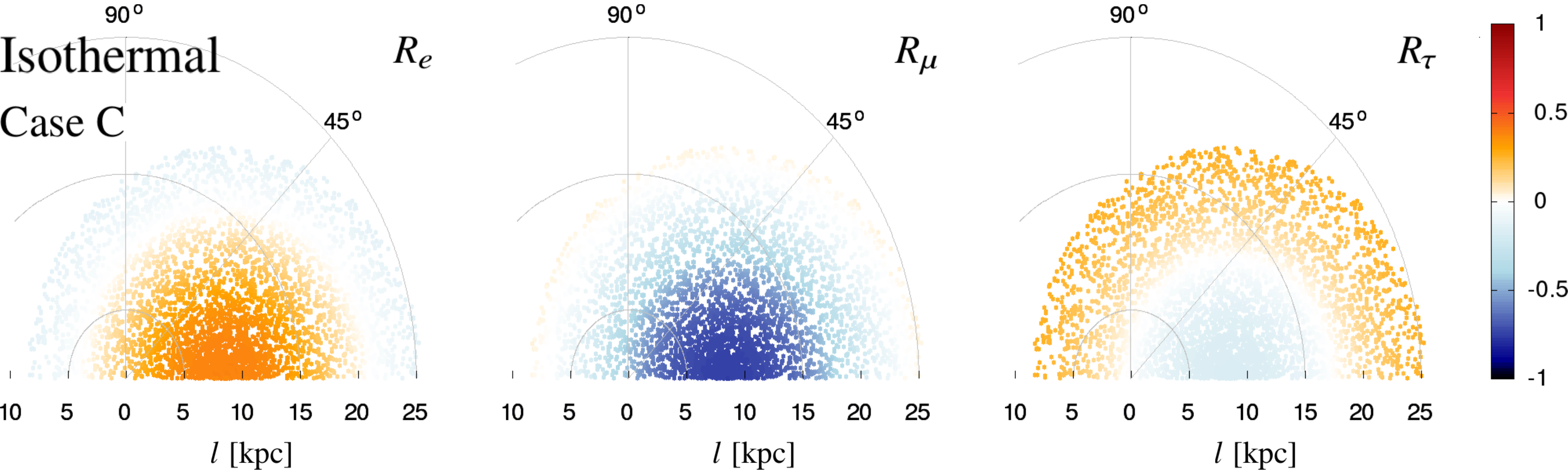}
\caption{\label{fig:circlesISO} $R_\beta$ maps for different neutrino production position. Same initial state as in Fig.~\ref{fig:circles} but DM distribution corresponds to an isothermal profile. Panels from left to right correspond to $R_e$, $R_\mu$, $R_\tau$ respectively and all plots are for the benchmark case C.}
\end{figure*}

In Fig.~\ref{fig:circlesISO} we present the $R_{\beta}$ map using the benchmark C and the isothermal profile.
The direct comparison of Figs.~\ref{fig:circles} and~\ref{fig:circlesISO} shows that the flavor conversion is enhanced towards the galactic center.
One can notice that, thanks to the inner cusp, the NFW case is rather different from the isothermal one only in the $\sim 1$~kpc region surrounding the GC.
This promising feature might indicate that a DM--MSW effect can reveal relevant information about the innermost GC.
We also observe that the obtained values of $R_{\beta}$ are symmetrical with respect to the GC.

We show in Fig.~\ref{fig:Triang120} the predicted flavor compositions at Earth for neutrinos originally produced as (1:2:0) and (1:0:0).
As initial neutrino position we have considered all points inside a 20 kpc radius from the GC.
In the left panel, we compare the overall region for a constant potential (reddish area) with the region resulting from the NFW profile (blueish region). Also, we include the uncertainty bands as well as the best fit point obtained from the IceCube results on neutrino flavor composition~\cite{2015arXiv151005223T}. 
The values of all entries of the potential $|V_{ij}^\oplus|$  range from $10^{-24}$~eV to $10^{-16}$~eV.
The main difference between the blueish and reddish areas arises because the DM distributions are different near the GC and in the outskirts of the galaxy.
The strongest impact of the DM interaction on neutrino oscillations on these production zones broadens the blueish area beyond the expected region for a constant potential.

The right panel of Fig.~\ref{fig:Triang120} shows the possible flavor compositions produced from an initial flavor (1:0:0) and a NFW DM profile with different values of $V_{ij}^{\oplus}$.
Each colored area arises as the result of a scan with a maximum value of $|V_{ij}^{\oplus}|$.
We observe that  values of $V_{ij}^{\oplus}$ as larger as $10^{-17}$~eV can almost cover the full flavor triangle.
However, as soon as the maximum value of $V_{ij}^{\oplus}$ is reduced, the obtained areas tend to converge to the vacuum solution.
Maximum values of $V_{ij}^{\oplus}$ smaller than $10^{-21}$~eV might not be enough to explain the 3$\sigma$ region of IceCube.
We observe that,  for the same range of $V_{ij}^{\oplus}$, the area produced by a homogeneous dark matter profile (Fig.~\ref{fig:triancte}) is much more restricted than the one obtained  from a NFW profile.

The best fit point and allowed bands obtained from the latest IceCube data are qualitatively compatible with the constant potential and the DM profile explanations\footnote{A more detailed analysis including full energy range and cuts will be required in order to check the agreement in a quantitative way.}.
Nevertheless, we note that for the constant DM case the values of the potential $V_{ij}^{\oplus}$ required to explain the observed flavor composition in IceCube are larger than the ones in the DM profile case.
This is due to the effect of the variable DM profile density over neutrino oscillations. 
The change in the effective potential felt by neutrinos will depend on their arrival direction, e.g. neutrinos coming from regions near the GC would feel larger modifications in their effective potential.
This indicates that flavor neutrino composition might be  angularly correlated and then, it may depend on the part of the sky accessible to IceCube.\\

\section{Discussion}
\label{sec:disc}

In the previous sections we have analyzed the effect of DM on neutrino oscillations
using an effective potential parametrized in Eq.(\ref{eq:effpot36}).
%
%
In a more generic framework, considering New Physics beyond the Standard Model, the interaction between both particles can also be expressed as
\begin{equation}\label{eq:effpotprime}
V_{ij} = \lambda^{\prime}_{ij} \, G^{\prime}_F \, \frac{\rho_{\rm DM}}{m_{\rm DM}} \, .
\end{equation}
This equation allows the reinterpretation of the effective potential in terms of a new interaction strength $G^{\prime}_F$ as well as on the DM mass, $m_{\rm DM}$.
We highlight that Eqs.~(\ref{eq:effpot36}) and (\ref{eq:effpotprime}) are parameterizations of the potential where the structure of the $\lambda_{ij} (\lambda_{ij}')$ parameters will depend on the  choice of a given particle physics model.
For simplicity, we will assume that the interaction between neutrinos and DM particles happens via the interchange of a $Z^{\prime}$-like boson. In this case, the \emph{primed} Fermi constant is related to the standard $G_F$ by
\begin{equation}
G^{\prime}_F = \frac{m^2_{Z}}{m^2_{Z^{\prime}}} \, G_F \, ,
\end{equation} 
with $m_{Z} \simeq 91$~GeV. Here, the mass of the mediator and the interaction strength are tightly related.
This scaling is valid only for the coherent scattering regime if there is no momentum transferred by the mediator, or if the mediator is so heavy that it can be integrated out.\\

It is important to mention that, besides the effect of forward coherent scattering encoded in the DM potential $V$,  neutrinos might actually scatter on the DM halo.
This could disrupt the effect of neutrino oscillations, modifying also their arrival directions and energy spectrum~\cite{Davis:2015rza}.
To this end, we have to ensure that the mean free path due to the neutrino-DM scattering cross section at Earth,
\begin{equation}
l_{\nu}=\left(\sigma_{\nu \chi} \frac{\rho_{\rm DM}}{m_{\rm DM}}\right)^{-1}=\left(\frac{\sigma_{\nu\chi}}{8.1\times10^{-22}{\rm cm}^2}\right)^{-1} \left(\frac{m_{\rm DM}}{\rm GeV}\right) {\rm kpc},
\end{equation}
is large enough, allowing neutrinos to cross the galaxy and being affected only at the level of oscillations.
%

\begin{figure*}[t]
\centering
\includegraphics[width=0.49\textwidth]{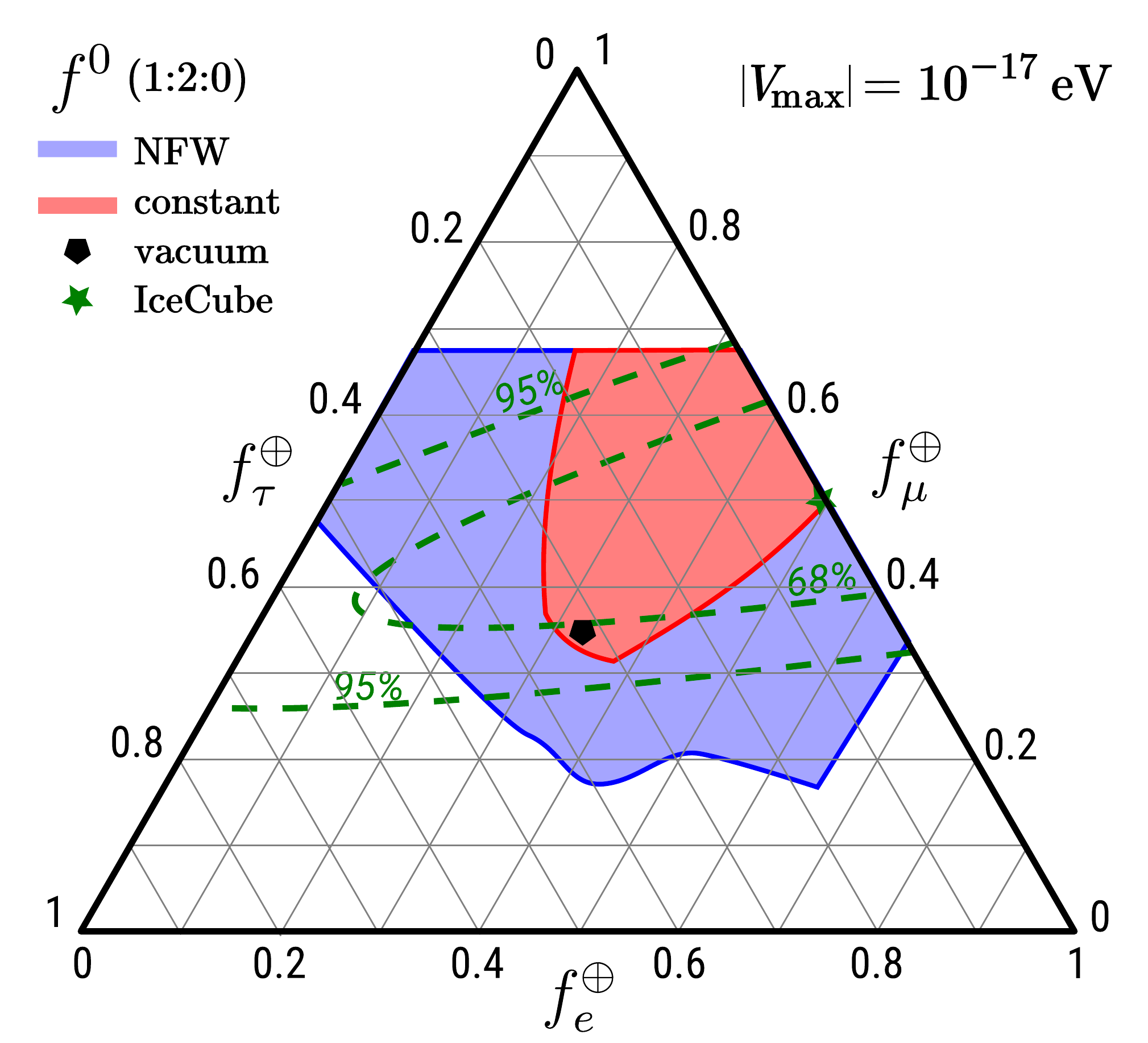}
\includegraphics[width=0.49\textwidth]{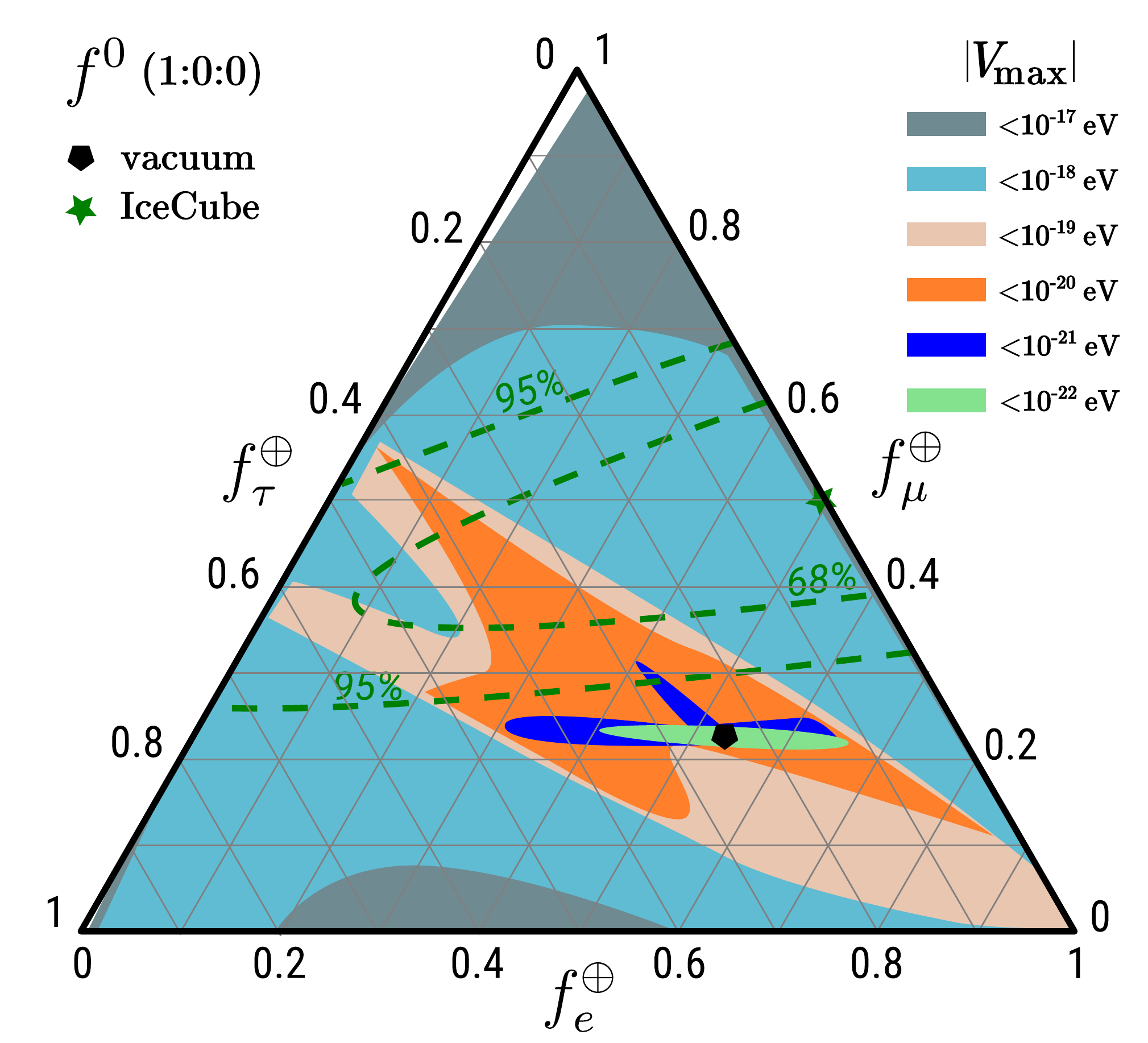}
\caption{\label{fig:Triang120} Flavor triangles for the initial states (1:2:0) (left) and (1:0:0) (right). Left panel shows the flavor area covered by a homogeneous DM profile (reddish area) and the one covered by the effect of a NFW profile (blueish area). Right panel shows the areas covered by imposing a maximum value for $V_{ij}^{\oplus}$ and a NFW profile. We observe that for smaller maximum values the area is closer to the solution in vacuum. The best fit point and 68\% and 95\% C.L. allowed regions from IceCube data are also shown.}
\end{figure*}

For the case of galactic neutrinos, we set $l_{\nu}$ at Earth to be 50~kpc which corresponds to a cross section $\sigma_{\nu \chi} = 1.62 \times 10^{-23} \,  \left(m_{\rm DM}/{\rm GeV}\right) {\rm cm}^2$.
This value guarantees that, apart from the effect of coherent forward scattering, neutrinos would rarely scatter along any trajectory, including the ones passing through the GC for a NFW profile.
A complementary study for ultra-light DM, in the regime where neutrino-DM scattering cross section has a relevant role for the neutrino propagation 
 is analyzed in Ref.~\cite{Reynoso:2016hjr}.

In a more extreme case, we also consider the bound $\sigma_{\nu \chi} < 10^{-33} \,  \left(m_{\rm DM}/{\rm GeV}\right) {\rm cm}^2$, which comes from the CMB analysis when DM-neutrino interactions are allowed~\cite{Wilkinson:2014ksa}.
This bound corresponds to $l_{\nu} > 10^6$~Gpc at Earth.
Let us remark that this bound applies for DM-neutrino cross sections at the MeV scale and therefore its value might be different for the neutrino energies considered in this work, depending on the particle physics model considered.
In what follows, we will use these two values of the neutrino mean free path as benchmarks to discuss the dependence of the effective potential on the remaining parameters, $\lambda^{\prime}$, $G_{F}^{\prime}$ and $m_{Z^{\prime}}$.
We will also consider a particle physics scenario with unconstrained mean free path in order to better understand the role of the parameters.\\

In Table~\ref{tab:interpr2}, we present six different choices for the involved parameters that can reproduce the three selected values for the effective potential at Earth.
These values have been chosen in the range from $10^{-21}$ to $10^{-17}$~eV, a representative range of the potential that might produce sizable effects on the neutrino oscillation flavor content as it is shown in Fig.~\ref{fig:Triang120}.
All over the table we have considered that  the neutrino-DM  cross section takes the form
\begin{equation}
\sigma \propto {\lambda^{\prime}}^2 G_F^{\prime} \simeq {\lambda^{\prime}}^2 \, \left( \frac{m_{Z^{\prime}}}{\rm GeV} \right)^{-2} \, 3.75 \times 10^{-29} \, {\rm cm}^2\,,
\end{equation}
well motivated for the case \mbox{$\sqrt{s} \gg m_{Z^{\prime}}, m_{\rm DM}$}.
At this point,  the functional relation between $V^{\oplus}$, $l_\nu$ and the rest of the parameters ($m_{\rm DM}$, $\lambda^\prime$ and $G_F^\prime$) allows us to describe the effect on neutrino oscillations by fixing only three of them.\\

The scenarios described in the table are the following:

\begin{itemize}
\item {\bf Weak scale}.- We assume for all cases that $G_{F}^{\prime} = G_F$.
If we further impose $\lambda'=1$ (case {\bf a}), the DM mass has to lie in the mass range from $10^{-12}$ to $10^{-8}$~eV.
This corresponds to an extremely light DM particle which is in the spirit of Axion Bose-Einstein Condensates (BEC)~\cite{2009PhRvL.103k1301S}, although the mass we obtain is (at best) two orders of magnitude smaller than the BEC case.
One can find models with such extremely light candidates like in scalar (wave) dark matter models~\cite{Matos:2000ss}.
As the result of the chosen $\lambda'$ and $G_{F}^{\prime}$, the neutrino mean free path is in the sub--pc range. 
This small value of $l_{\nu}$ might imply that high energy neutrinos are screened due to their interaction with the DM halo~\cite{Davis:2015rza}.
If we now impose $l_\nu = 50$~kpc (case {\bf b}), as discussed above, the values of $\lambda'$ are reduced to the range between $10^{-7}$ and $10^{-11}$,
and the DM mass becomes even lighter, laying on the range from  $10^{-15}$ to $10^{-23}$ eV.

\item {\bf 100 GeV DM}.-
Here we fix $l_\nu$ at Earth to the two values previously discussed.
For $l_\nu = 50$~kpc (case {\bf a}), we obtain the same values of $\lambda'$ as before.
In this case, $m_{Z^{\prime}}$ takes sub-eV values, $10^{-2}$ to $10^{-6}$~eV, which 
may be in the line of models with light mediators in the neutrino sector~\cite{2015arXiv151002201F}.
On the other hand, for $l_{\nu} = 10^6$~Gpc (case {\bf b}), the values of $\lambda'$ are in the range from $10^{-17}$ to $10^{-21}$, significantly smaller than for larger values of $l_{\nu}$.
The same happens for $m_{Z^{\prime}}$ that now takes values within the range from $10^{-7}$ to $10^{-11}$~eV, well below the predictions of case (a), 
In general terms, larger values of the mean free path imply lower values of $\lambda'$ and $m_{Z^{\prime}}$.

\item {\bf 1 keV DM}.- Here, the lower value of the DM mass results in an increase value of the number density of DM particles,  $N_{\chi}$.
Depending on the value of $V^{\oplus}$, this could lead to a mediator mass of the order of few eV,  in agreement with models including light mediators.
The values of $\lambda'$ remain unchanged with respect to the 100 GeV DM case. 
\end{itemize}

\begin{table*}[!t]
\centering
\begin{tabular}{|c|ccc|}
\hline
\phantom{\LARGE{MM}}$V_{11}^{\oplus}$ [eV] \phantom{\LARGE{MM}} & \phantom{\LARGE{M}} $10^{-21}$ \phantom{\LARGE{M}}&\phantom{\LARGE{M}} $10^{-19}$ \phantom{\LARGE{M}}&\phantom{\LARGE{M}} $10^{-17}$ \phantom{\LARGE{M}}\\
\hline\hline
\multicolumn{1}{|r}{{\bf Weak scale (a)} assumptions:} & \multicolumn{3}{l|}{$G_F^{\prime} = G_F$, $\lambda_{11} = 1$}  \\
\hline
$m_{\mathrm{DM}}$ [eV] & $10^{-8}$ & $10^{-10}$ & $10^{-12}$\\
$l_{\nu}$ [pc]         & $10^{-2}$ & $10^{-4}$  & $10^{-6}$ \\
\hline\hline
\multicolumn{1}{|r}{{\bf Weak scale (b)} assumptions:} & \multicolumn{3}{l|}{$G_F^{\prime} = G_F$, $l_{\nu} = 50$ kpc}  \\
\hline
$\lambda_{11}$ & $10^{-7}$ & $10^{-9}$ & $10^{-11}$ \\
$m_{\mathrm{DM}}$ [eV] & $10^{-15}$ & $10^{-19}$ & $10^{-23}$\\
\hline\hline
\multicolumn{1}{|r}{{\bf 100 GeV DM (a)} assumptions:} & \multicolumn{3}{l|}{$m_{\mathrm{DM}} = 100$ GeV, $l_{\nu} = 50$ kpc}  \\
\hline
$\lambda_{11}$ & $10^{-7}$& $10^{-9}$& $10^{-11}$ \\
$m_{Z'}$ [eV]  & $10^{-2}$& $10^{-4}$& $10^{-6}$ \\
\hline\hline
\multicolumn{1}{|r}{{\bf 100 GeV DM (b)} assumptions:} & \multicolumn{3}{l|}{$m_{\mathrm{DM}} = 100$ GeV, $l_{\nu} = 10^{6}$ Gpc}  \\
\hline
$\lambda_{11}$ & $10^{-17}$ & $10^{-19}$ & $10^{-21}$ \\
$m_{Z'}$ [eV]  & $10^{-7}$  & $10^{-9}$  & $10^{-11}$ \\
\hline\hline
\multicolumn{1}{|r}{{\bf 1 keV DM (a)} assumptions:} & \multicolumn{3}{l|}{$m_{\mathrm{DM}} = 1$ keV, $l_{\nu} = 50$ kpc} \\
\hline
$\lambda_{11}$ & $10^{-7}$ & $10^{-9}$ & $10^{-11}$ \\
$m_{Z'}$ [eV]  & $10^{2}$  & $1$       & $10^{-2}$ \\
\hline\hline
\multicolumn{1}{|r}{{\bf 1 keV DM (b)} assumptions:} & \multicolumn{3}{l|}{$m_{\mathrm{DM}} = 1$ keV, $l_{\nu} = 10^{6}$ Gpc}  \\
\hline
$\lambda_{11}$ & $10^{-17}$ & $10^{-19}$ & $10^{-21}$\\
$m_{Z'}$ [eV]  & $10^{-3}$  & $10^{-5}$  & $10^{-7}$ \\
\hline
\end{tabular}
\caption{
\label{tab:interpr2}
Particle physics interpretations for three given values of the effective potential at Earth, $V_{11}^{\oplus}$.
Six different scenarios are displayed, with the corresponding assumptions  shown in the upper row of each block.
For each case, second, third and fourth columns show the values of the remaining parameters for the corresponding values of $V_{11}^{\oplus}$.
Implications of each case are described in the text.}

\end{table*}

From the model-building point of view, this type of interaction presents tantalizing insights between DM and neutrinos.
For instance, models connecting neutrino masses with dark matter candidates might produce flavor ratios at Earth outside the expectations in vacuum. 
These particle model scenarios might imply interesting consequences and motivate further analyses.

All over our calculations, we have parameterized the DM-neutrino interaction in such a way that DM interacts with both neutrinos and antineutrinos.
Let us remark that the interaction with neutrinos and antineutrinos is intrinsically related to the nature of DM and the particle mediator.
For instance, in models of asymmetric DM, we might expect that DM affects only neutrinos or antineutrinos.
On the other hand, this could also happen in symmetric DM scenarios (WIMP-like), where the nature of the mediator might be responsible for the non-symmetric effect, e.g. coupling DM to antineutrinos and anti-DM to neutrinos.
Let us highlight that this asymmetry is already present in the standard MSW effect, since most of the media are mainly composed by matter and no antimatter.
In a more general scenario, the predicted neutrino flavor composition at Earth would be different for neutrinos and antineutrinos.
%

Another aspect to be considered is the relation between the neutrino arrival direction and the observed flavor composition.
The DM distribution molds the observed  neutrino flavor composition depending on its  production point within the Milky Way, as seen in Subsection~\ref{ssec:oscdist}.
This implies that the location of the neutrino observatory plays an important role on the observation of the neutrino flavor content.
For instance, KM3NeT and IceCube  will have access to different parts of the sky.
One will be able to observe neutrinos coming from directions towards the galactic center while the other one may observe neutrinos coming from the Milky Way outskirts.
In consequence, the averaged flavor composition observed at each experiment may be different.
Therefore, it will be important to include the arrival direction information in future analyses in order to disentangle a possible correlation between neutrino flavor and DM distribution.\\

Although it has not been discussed in this work,  the effect of DM on neutrino oscillations might also be relevant for extragalactic neutrinos.
In consequence, PeV neutrinos coming from regions above or below the galactic plane can also present different flavor compositions with respect to the expectations in vacuum.
In this case, the effect would depend on the properties of the DM halo where the neutrino source is located and on how it compares with the one of the Milky Way.
Besides that, a larger $l_\nu$ might be required, depending on the size and density of the departing DM halo.
The distance to the source would also be relevant in this case, due to the loss of coherence in the neutrino propagation.\\
%

\section{Conclusions}
\label{sec:conc}

The observation of PeV neutrinos opens a new window to explore astrophysical processes and neutrino sources.
Independently of the source location, neutrinos must travel across the galactic DM halo.
Depending on the  distance to the source, we can observe deviations with respect to neutrino oscillations in vacuum due to the interaction between DM and neutrinos.
In this work, we have assumed the most generic interaction between dark matter and neutrinos, as described in Eq.~(\ref{eq:effpot36}).
We have studied the effects of a uniform effective potential by assuming a homogeneous DM halo.
We have obtained that the flavor composition at Earth is rather different with respect to the expected composition in vacuum for values of the effective potential larger than $10^{-20}$~eV.
In the analysis, we have used different initial flavor compositions inspired by astrophysical processes. 
We have scanned on a range of effective potential values $V_{ij}$  between $10^{-23}$ and $10^{-16}\, {\rm eV}$, finding different  covered areas in the flavor triangle.
We have checked that these areas can not be enlarged by considering larger values of the interaction potential.
In particular, we have found that the full triangle is not accessible for the case of homogeneous DM and a fixed type of neutrino source.
Along this work, we have also checked that our results hold for both, normal and inverted hierarchy.\\

In a more realistic scenario, we have included a radial density distribution for the DM halo of the Milky Way by using NFW and isothermal DM profiles.
As a consequence, the effective potential acquires a spatial dependence and therefore  the neutrino propagation equation has to be solved numerically taken into account the DM distribution profile\footnote{In the case of adiabatic neutrino propagation, however, this calculation gets simplified and only the DM density at the initial and final points of the neutrino path are relevant.}.
In addition, we have considered that neutrinos are produced homogeneously on the whole galactic plane up to 20 kpc from the GC.
The covered flavor composition areas depend on the considered DM distribution and we found that
the more realistic case with a  DM profile produces a larger area in the flavor triangle than the homogeneous case. 
This implies a richer scenario than other New Physics cases (see Ref.~\cite{2015PhRvL.115p1302B} and references within), e.g. violation of Lorentz symmetry~\cite{Arguelles:2015dca}.

This effect can also be used to probe particle physics models beyond the SM for DM-neutrino interactions.
We provide different interpretations in terms of three simple scenarios: {\bf weak scale}, implying an extremely light DM with mass $\mathcal{O}(10^{-23} - 10^{-15}) \, {\rm eV}$; {\bf 100 GeV} and {\bf 1 keV DM} cases. The latter two cases can be explained in terms of new very light mediators with masses \mbox{$\mathcal{O}($sub-eV $-$ keV$)$.}
Each of these scenarios might provide very interesting insights in relation to models of neutrino masses~\cite{2008PhRvD..77d3516B}.

A more detailed analysis considering the neutrino arrival direction might further reveal the presence of the effect considered in this work.
For instance, depending on the available sky for observation, the averaged neutrino flavor content might be different.
This could be the case of IceCube and KM3NeT.
For the timescale of the IceCube observatory, the most optimistic expectations indicate that 10 years of data taking would be enough to improve the estimation of the neutrino flavor composition at Earth~\cite{2015PhRvL.115p1302B,2016PhRvD..93h5004S}.
This would happen within the timescale of future extensions of IceCube, as the proposed IceCube-Gen2 detector~\cite{2015arXiv151005228T}.
In this scenario, our hypothesis could be tested by imposing constraints on the maximum value of the effective potential due to the DM-neutrino interaction.
During this period, KM3NeT will be in operation as well~\cite{Adrian-Martinez:2016fdl} and it hopefully will present its own results for the neutrino flavor composition at Earth. The comparison of the results of both observatories will allow us to test the hypothesis presented in this work in further detail.\\

\begin{acknowledgements}
We acknowledge \href{http://inspirehep.net/author/profile/O.G.Miranda.1}{O.~G.~Miranda}, \href{http://inspirehep.net/author/profile/S.Pastor.2}{S. Pastor}, \href{http://inspirehep.net/author/profile/B.Panes.1}{B.~Panes}, and \href{http://inspirehep.net/author/profile/N.Rojas.1}{N. Rojas} for useful discussions and comments.
This work was supported by the Spanish MINECO under grants FPA2014-58183-P, and MULTIDARK CSD2009-00064 (Consolider-Ingenio 2010 Programme); by Generalitat Valenciana grant PROMETEOII/2014/084, and Centro de Excelencia Severo Ochoa SEV-2014-0398.
\href{http://goo.gl/CHlgQR}{P.~FdS.} is supported by the FPU Scholarship FPU13/03729 (MINECO).
\href{http://goo.gl/00TnL}{R.~L.} is supported by a Juan de la Cierva contract JCI-2012-12901 (MINECO).
\href{http://goo.gl/x8CPgJ}{M.~T.} is supported by a Ram\'{o}n y Cajal contract (MINECO).
\end{acknowledgements}

\appendix

\section{Adiabaticity condition for  neutrino propagation}
\label{se:adia}

The equation of motion for the neutrino weak eigenstates in matter is given by~\cite{Wolfenstein:1977ue}

\begin{figure*}[!t]
\centering
\includegraphics[width=\figwidth]{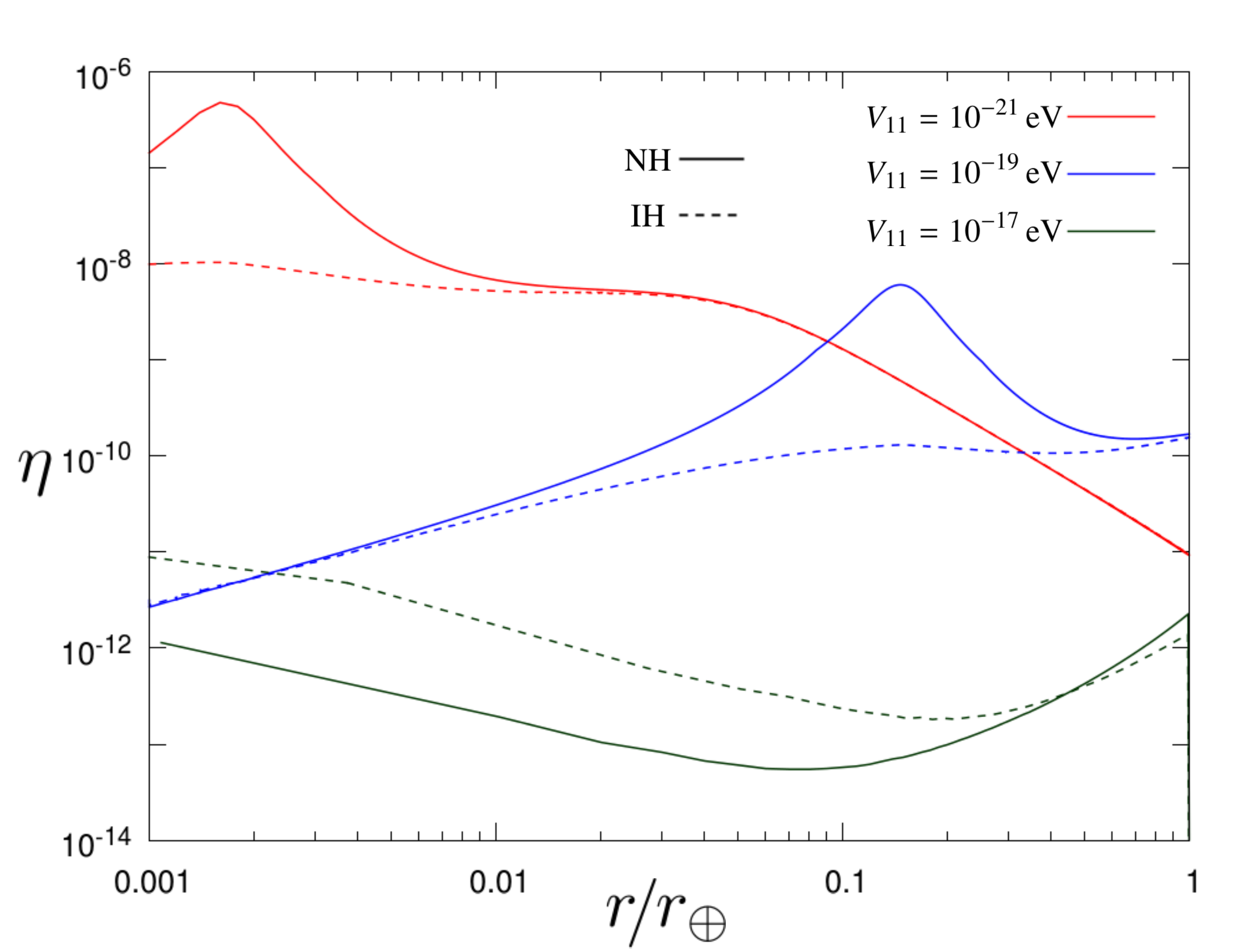}
\caption{\label{fig:adiabat} Adiabaticity parameter $\eta$ for a 1 PeV neutrino crossing the GC as a function of the distance from the GC,  $r$, given in units of the distance to the Earth, $r_\oplus$. A NFW DM profile is assumed. The variation of $\eta$ is shown for three different effective potentials at Earth, $V_{11}^{\oplus}$, assuming normal (solid line) and inverted (dashed line) neutrino mass hierarchy.}
\end{figure*}

\begin{equation}
i\,\dot{\nu}_W = \mathcal{H}_{\rm tot} \nu_W,
\end{equation}
where $\nu_W$ is a vector containing the different neutrino flavour states, and $\mathcal{H}_{\rm tot}$ is the total hamiltonian. 
The relation between weak and mass eigenstates is given by
\begin{equation}
\nu_W = U \nu_m \, ,
\end{equation}
where $U$ is a unitary matrix that diagonalizes the total hamiltonian,
\begin{equation}
\mathcal{D} = U^\dagger \mathcal{H}_{\mathrm{tot}} U \, ,
\end{equation}
and $\nu_m$ is the vector containing the  effective neutrino mass eigenstates.
If neutrinos travel through a medium with varying effective potential,  the mixing matrix $U$ will vary along the neutrino path and, therefore, the equation of motion for the neutrino mass eigenstates will be
\begin{equation}
i \, \dot{\nu}_m = \left( \mathcal{D} - i U^\dagger \dot{U} \right) \nu_m \, ,
\label{eq:adiab}
\end{equation}
where the term $U^\dagger \dot{U}$ vanishes for a constant effective potential. 
However, whenever neutrinos travel through a varying density medium, this term is different from zero and $\nu_m$  no longer describes the mass eigenstates.
Nevertheless, for slow changing medium density (with respect to oscillation wavelength), $\nu_m$ can be approximated by true mass eigenstates. In this case the process is said to be \emph{adiabatic}.
On the contrary, in the non-adiabatic regime, where the medium density changes fast enough, there is a probability of level crossing between neutrino mass states known as Landau-Zener probability (See Refs.~\cite{landau1932, zener1932}).

To check that adiabaticity is fulfilled along the neutrino path, we define the adiabaticity parameter as
\begin{equation}
\eta = \frac{ \langle \left| U^\dagger \dot{U}  \right| \rangle}{\langle \left| \mathcal{D} - \mathrm{Tr}\left( \mathcal{D} \right)/3 \, \mathbb{I} \right| \rangle} \, ,
\end{equation}
where $\langle \cdots \rangle$ stands for the average over the matrix indices.
Notice that the diagonal elements of $U^\dagger \dot{U}$ are zero and the denominator is traceless.
Since $U^\dagger \dot{U}$ has only non-diagonal elements different from zero and $\mathcal{D}$ is a diagonal matrix, the non-diagonal elements of ($\mathcal{D} - i U^\dagger \dot{U}$) in Eq.~(\ref{eq:adiab}) come only from the second term.
Hence, a process is adiabatic if $\eta \ll 1$.
This condition ensures no mixing between the effective mass eigenstates induced by the variation of the medium and, therefore, $\nu_m$ stay as the true energy eigenstates of the hamiltonian along the neutrino path.
In consequence, one can evaluate the final neutrino flavor composition by considering only the effective DM potential at the beginning and end of the neutrino trajectory.

In Fig.~\ref{fig:adiabat}, we show the adiabaticity parameter for a 1 PeV neutrino crossing the GC for the DM-neutrino interaction cases presented in Tab.~\ref{tab:interpr2} and a NFW profile. The case with $V_{11}^{\oplus}=10^{-17}\;{\rm eV}$ corresponds to an effective potential that saturates neutrino oscillations. Therefore, the value of $\eta$ is very small along the neutrino path. The other values of the potential shown in the figure, $V_{11}^{\oplus}=10^{-19}\;{\rm eV}$ and $V_{11}^{\oplus}=10^{-21}\;{\rm eV}$, do not saturate oscillations at Earth but they do it instead at the very center of the galaxy, due to the larger DM density. This explains the peaks in $\eta$ followed by a decrease in the function when $r$ approximates to zero. The peaks correspond to regions with maximum change in the potential previous to reach the saturation values, i.e. the regions where the effective oscillation parameters are more sensitive to variations in the DM potential.
For a further discussion about adiabaticity see e.g. Section~8.3 in~\cite{kim1993neutrinos}, where the adiabaticity condition is studied under the two flavor neutrino oscillation approximation.


\def\apj{Astrophys.~J.}                       
\def\apjl{Astrophys.~J.~Lett.}                
\def\apjs{Astrophys.~J.~Suppl.~Ser.}          
\def\aap{Astron.~\&~Astrophys.}               
\def\aj{Astron.~J.}                           %
\def\araa{Ann.~Rev.~Astron.~Astrophys.}       %
\def\mnras{Mon.~Not.~R.~Astron.~Soc.}         %
\def\physrep{Phys.~Rept.}                     %
\def\jcap{J.~Cosmology~Astropart.~Phys.}      
\def\jhep{J.~High~Ener.~Phys.}                
\def\prl{Phys.~Rev.~Lett.}                    
\def\prd{Phys.~Rev.~D}                        
\def\nphysa{Nucl.~Phys.~A}                    

\bibliographystyle{apsrev4-1}
\bibliography{dm-nu-osc.bib}

\end{document}